\documentclass[12pt,a4wide]{article}
\usepackage{amsmath,amssymb,amsfonts}

\newcommand{\eqa}{\begin{eqnarray}}
\newcommand{\ena}{\end{eqnarray}}
\newcommand{\topstar}[1]{\setlength{\unitlength}{1mm}
\begin{picture}(2,0)(-1,-1.4)
   \put(0,0){\makebox(0,0){$#1$}}
   \put(0,2.4){\makebox(0,0){\mbox{\tiny$\star$}}}
\end{picture}}

\begin{document}
\begin{center}
{\large {\bf Fully Extended Quasi-Metric Gravity}}
\footnote{\*{Dedicated to the memory of my mother}}
\end{center}
\begin{center}
Dag {\O}stvang \\
{\em Department of Physics, Norwegian University of Science and Technology
(NTNU) \\
N-7491 Trondheim, Norway}
\end{center}
\begin{abstract}
The original theory of quasi-metric gravity, admitting only a partial coupling 
between space-time geometry and the active stress-energy tensor, is too 
restricted to allow the existence of gravitational waves in vacuum. Therefore, 
said theory can at best be regarded as a waveless approximation theory. 
However, the requirement that the weak-field limit of the contracted Bianchi
identities should be consistent with the Newtonian limit of the local 
conservation laws, forbids a full coupling between space-time geometry and the 
active stress-energy tensor. Nevertheless, in this paper it is shown how it is 
possible to relax the restrictions on quasi-metric space-time geometry 
sufficiently to avoid these problems. That is, the original quasi-metric field 
equations can be extended with one extra field equation, without having said 
full coupling and such that the contracted Bianchi identities have a sensible
Newtonian limit. For weak fields in vacuum, said extra field equation has a 
dynamical structure somewhat similar to that of its counterpart in canonical
general relativity (GR). In this way, the prediction of weak GR-like 
gravitational waves in vacuum becomes possible. Moreover, exact results from 
the original quasi-metric gravitational theory are recovered for metrically 
static systems and for isotropic cosmology. This means that the current 
experimental status of the extended quasi-metric gravitational theory is the 
same as for the original theory, except for the prediction of weak GR-like 
gravitational waves in vacuum.
\\
\end{abstract}
\topmargin 0pt
\oddsidemargin 5mm
PACS: 04.50.Kd, 04.90.+e   
\newpage
\section{Introduction}
The so-called quasi-metric framework (QMF), an alternative geometric framework 
for formulating relativistic gravitation, was invented a number of years ago 
[1, 2]. Currently, the QMF turns out to have a non-viable status due to its
predicted properties of the cosmic relic neutrino background (assuming 
standard neutrino physics) [3]. This status could possibly change if future
experiments show evidence of the necessary non-standard neutrino physics needed
to resolve the apparent conflict with observations. Now said predicted 
properties of the cosmic neutrino background depend crucially on the neutrino 
physics, but not on the gravitational sector of the QMF. Therefore, the 
currently non-viable status of the QMF is valid independently of any particular
theory of quasi-metric gravity.

However, the original theory of quasi-metric gravity (OQG) has a much more 
serious problem. That is, due to the very restricted form postulated for
quasi-metric space-time geometry, only a partial coupling is possible between 
the active stress-energy tensor and space-time curvature. Unfortunately, this
makes the original theory essentially waveless since said restricted form
of the quasi-metric space-time geometry makes it fully determinable by the 
matter sources alone. Unlike general relativity (GR), where the field equations
directly determine the Ricci tensor only, leaving the Weyl curvature free,
the OQG leaves free no aspect of quasi-metric space-time geometry. Thus no 
kinds of gravitational waves in vacuum can exist according to the OQG. With 
the recent direct experimental evidence for GR-like gravitational waves, this 
means that the OQG must be abandoned and at best be treated as a waveless 
approximation theory.

Now it turns out that it is not possible to have both a sensible weak-field
limit of the contracted Bianchi identities and in addition allowing a full 
coupling of quasi-metric space-time geometry to the active stress-energy tensor.
In fact, if one tries this, the weak field limit of the contracted Bianchi 
identities will be inconsistent with that of the local conservations laws. 
Fortunately, it is possible to avoid this problem by relaxing the postulated 
form of the quasi-metric space-time geometry sufficiently to allow the 
existence of one extra field equation which is only partially coupled to the 
active stress-energy tensor. As we shall see, for weak gravitational fields in 
vacum, this extra field equation has, by construction, a dynamical structure 
somewhat similar to its counterpart in canonical GR. That is, due to some 
resemblance in form to the dynamical structure of canonical GR, the extended 
field equations predict weak GR-like gravitational waves in vacuum. This means 
that there is some hope that the extended version of quasi-metric gravity may 
eventually turn out to be viable.
\section{Fully extended quasi-metric gravity}
\subsection{General space-time geometry}
The QMF has been published in detail in [1] (see also [2]). Here we include 
only the minimum basics and the required adoptions made to acommodate the 
extended quasi-metric field equations.

In short, the basic theoretical motivation for introducing the QMF is to 
eliminate the in principle enormous number of potential possibilites regarding 
cosmological dynamics and evolution existing for metric theories of gravity. 
Since the Universe is presumely unique, the existence of such multiple 
potential possibilities is a problem and a liability because at most one of 
said possibilities is available for testing. This means that the predictive 
power of a cosmology based on metric theories of gravity is in general weak; 
either one must try to solve said problem in some {\em ad hoc} manner, or one 
is at best limited to fitting a number of cosmological parameters in a 
consistent way. On the other hand the QMF solves said problem in a geometrical
way; in quasi-metric space-time there is no room at all for potential 
cosmological dynamics involving the cosmic expansion. This follows from the 
idea that the cosmic expansion should be described as a general phenomenon not 
depending on the causal structure associated with any pseudo-Riemannian 
manifold. In other words, within the QMF, the cosmic expansion has nothing to 
do with causality or dynamics; rather it is an ``absolute'' intrinsic property 
of quasi-metric space-time itself. And as we shall see in this section, the 
geometrical structure of quasi-metric space-time ensures that this alternative 
way of describing the cosmic expansion is mathematically consistent and 
fundamentally different from its counterpart in GR. In what follows it is 
shown how said motivation is realized geometrically.

The geometrical basis of the QMF consists of a 5-dimensional 
differentiable manifold with topology ${\cal M}{\times}{\bf R}_1$, where 
${\cal M}={\cal S}{\times}{\bf R}_2$ is a Lorentzian space-time manifold, 
${\bf R}_1$ and ${\bf R}_2$ are two copies of the real line and ${\cal S}$ is a 
compact 3-dimensional manifold (without boundaries). That is, in addition to 
the usual time dimension and 3 space dimensions, there is an extra time 
dimension represented by {\em the global time function} $t$ introduced as a
global time coordinate on ${\bf R}_1$. The reason for introducing this extra 
time dimension is that by definition, $t$ parametrizes any change in the 
space-time geometry that has to do with the cosmic expansion. By construction, 
the extra time dimension is degenerate to ensure that such changes will have 
nothing to to with causality. Mathematically, to fulfil this property, the 
manifold ${\cal M}{\times}{\bf R}_1$ is equipped with two degenerate 
5-dimensional metrics ${\bf {\bar g}}_t$ and ${\bf g}_t$. The metric 
${\bf {\bar g}}_t$ is found from field equations as a solution, whereas the 
``physical'' metric ${\bf g}_t$ can be constructed (locally) from 
${\bf {\bar g}}_t$ (for details, see refs. [1, 2]).

The global time function is unique in the sense that it splits quasi-metric 
space-time into a unique set of ``distinguished'' 3-dimensional spatial 
hypersurfaces called the {\em fundamental hypersurfaces (FHSs).} Observers 
always moving orthogonally to the FHSs are called {\em fundamental observers 
(FOs)}. The topology of ${\cal M}$ indicates that there also exists a unique 
``preferred'' ordinary global time coordinate $x^0$. We use this fact to 
construct the 4-dimensional quasi-metric space-time manifold $\cal N$ by 
slicing the submanifold determined by the equation $x^0=ct$ out of the 
5-dimensional differentiable manifold. (It is essential that this slicing is 
unique since the two global time coordinates should be physically equivalent; 
the only reason to separate between them is that they are designed to 
parametrize fundamentally different physical phenomena.) The general 
domain of applicability of the 5-dimensional degenerate metric fields 
${\bf {\bar g}}_t$ and ${\bf g}_t$ is limited to $\cal N$. Moreover, their 
degeneracy means that they may be regarded as one-parameter families of 
Lorentzian 4-metrics on $\cal N$. Note that there exists a set of particular 
coordinate systems especially well adapted to the geometrical structure of 
quasi-metric space-time, {\em the global time coordinate systems (GTCSs)}. A 
coordinate system is a GTCS iff the time coordinate $x^0$ is related to $t$ 
via the equation $x^0=ct$ in ${\cal N}$.

In what follows, we will use index notation where Greek indices are ordinary
space-time indices taking integer values in the range $0..3$, while Latin 
indices are space indices taking integer values in the range $1..3$. Any 
implicit dependence on $t$ will be indicated with a separate index, e.g., the 
family ${\bf {\bar g}}_t$ has the space-time coordinates 
${\bar g}_{(t){\mu}{\nu}}$. Moreover, Einstein's summation convention will be used
throughout. Using said notations, and expressed in a suitable GTCS, we now 
write down the most general form allowed for the family ${\bf {\bar g}}_t$ 
including both explicit and implicit dependences on $t$. That is, a general 
family ${\bf {\bar g}}_t$ can be represented by the family of line elements 
valid on the FHSs (this may be taken as a definition)
\eqa
{\overline {ds}}_t^2={\bar N}_t^2{\Big \{ }
[{\bar N}_{(t)}^k{\bar N}_{(t)}^s{\tilde h}_{(t)ks}-1](dx^0)^2+
2{\frac{t}{t_0}}{\bar N}_{(t)}^k{\tilde h}_{(t)ks}dx^sdx^0+
{\frac{t^2}{t_0^2}}{\tilde h}_{(t)ks}dx^kdx^s{\Big \} }.
\ena
Here, $t_0$ is some arbitrary reference epoch (usually chosen to be the present
epoch) setting the scale of the spatial coordinates, ${\bar N}_t$ is the family
of lapse functions of the FOs and ${\frac{t_0}{t}}{\bar N}^k_{(t)}$ are the 
components of the shift vector family of the FOs in 
$({\cal N},{\bf {\bar g}}_t)$. Also, $d{\bar{\sigma}}_t^2{\equiv}{\bar h}_{(t)ks}
dx^kdx^s{\equiv}{\frac{t^2}{t_0^2}}{\bar N}_t^2{\tilde h}_{(t)ks}dx^kdx^s$ is the 
spatial line element family corresponding to the metric family 
${\bf {\bar h}}_t$ intrinsic to the FHSs. 

In the OQG, the form of the line element family (1) was severely restricted by 
postulating that the line element spatial metric family 
$d{\tilde{\sigma}}_t^2{\equiv}{\tilde h}_{(t)ks}dx^kdx^s$ (corresponding to the 
metric family ${\bf {\tilde h}}_t$) of the FHSs must be set equal to the metric
$S_{ks}dx^kdx^s$ of the 3-sphere (with radius equal to $ct_0$). The reason for 
this restriction was to ensure the uniqueness of $t$ by requiring the FHSs to 
be compact [1]. However, this requirement inevitably leads to some form of 
prior 3-geometry. Then said restriction was also thought to prevent the 
possibility that the prior 3-geometry might interfere with the dynamics of 
${\bf {\bar g}}_t$. On the other hand, except for the explicit dependence on 
$t$, the form of equation (1) may seem completely general. But this is not 
really so since, as we shall see later, in order to have a potentially viable 
theory it is necessary that the Ricci curvature scalar family ${\tilde P}_t$, 
calculated from the metric family ${\bf {\tilde h}}_t$, should take a 
restricted form. This means that the FHSs are still required to be compact and 
that there still will be prior 3-geometry. The difference from the original 
theory is that in the revised theory, the prior 3-geometry is less restrictive 
and it will be indirectly implemented via a certain term in the extended field 
equation (20) (or (23)) below rather than as an explicit restricion of 
equation (1).

The families ${\bf {\bar g}}_t$ and ${\bf g}_t$ are related by the (local)
transformation ${\bf {\bar g}}_t{\rightarrow}{\bf g}_t$ as described in
[1, 2]. A general form for the family ${\bf g}_t$ is given by the family of 
line elements (using a GTCS)
\eqa
{ds}_t^2=[N_{(t)}^kN_{(t)}^s{\hat h}_{(t)ks}-N^2](dx^0)^2+
2{\frac{t}{t_0}}N_{(t)}^k{\hat h}_{(t)ks}dx^sdx^0+
{\frac{t^2}{t_0^2}}{\hat h}_{(t)ks}dx^kdx^s,
\ena
where the symbols have similar meanings to their (barred) counterparts in 
equation (1) (the counterpart to ${\bar h}_{(t)ks}$ is
$h_{(t)ks}{\equiv}{\frac{t^2}{t_0^2}}{\hat h}_{(t)ks}$). Note that the 
propagation of sources (and test particles) is calculated by using the 
equations of motion in $({\cal N},{\bf g}_t)$ (see equation (7) below). 
Moreover, since the proper time as measured along a world line of a FO should 
not directly depend on the cosmic expansion, the lapse function $N$ should not 
depend explicitly on $t$. Therefore, any potential $t$-dependence of $N$ must 
be eliminated by substituting $t$ with $x^0/c$ (using a GTCS) whenever it 
occurs before using the equations of motion. In the same way, any extra 
$t$-dependence of ${\bf g}_t$ coming from the transformation 
${\bf {\bar g}}_t{\rightarrow}{\bf g}_t$ must be eliminated. Consequently, any 
$t$-dependence of ${\hat h}_{(t)ks}$ will stem from that of ${\tilde h}_{(t)ks}$. 
Also note that, if for some reason one wants to use the equations of motion 
in $({\cal N},{\bf {\bar g}}_t)$, any explicit dependence of ${\bar N}_t$ on 
$t$ must be eliminated as well.

Next, $({\cal N},{\bf {\bar g}}_t)$ and $({\cal N},{\bf g}_t)$ are equipped 
with linear and symmetric connections ${\ }{\topstar{\bf {\bar {\nabla}}}}{\ }$
and ${\ }{\topstar{\bf {\nabla}}}{\ }$, respectively. These connections are
identified with the usual Levi-Civita connection for constant $t$, yielding the
standard form of the connection coefficients not containing $t$.
The rest of the connection coefficients are determined by the condition that,
the connections ${\ }{\topstar{\bf {\bar {\nabla}}}}{\ }$
and ${\ }{\topstar{\bf {\nabla}}}{\ }$ should be compatible with the 
non-degenerate part of ${\bf {\bar g}}_t$ and ${\bf g}_t$, respectively. That 
is, we have the conditions
\eqa
{\topstar{\bf {\bar {\nabla}}}}_{\frac{\partial}{{\partial}t}}
{\bf {\bar g}}_t=0, \qquad
{\topstar{\bf {\bar {\nabla}}}}_{\frac{\partial}{{\partial}t}}
{\bf {\bar n}}_t=0, \qquad
{\topstar{\bf {\nabla}}}_{\frac{\partial}{{\partial}t}}
{\bf g}_t=0, \qquad
{\topstar{\bf {\nabla}}}_{\frac{\partial}{{\partial}t}}
{\bf n}_t=0,
\ena
where ${\bf {\bar n}}_t$ and ${\bf n}_t$ are families of unit normal vector 
fields to the FHSs in $({\cal N},{\bf {\bar g}}_t)$ and $({\cal N},{\bf g}_t)$,
respectively. The conditions shown in equation (3) will hold if we make the 
requirements (where a comma denotes taking a partial derivative)
\eqa 
{\frac{\partial}{{\partial}t}}{\Big [}{\bar N}_{(t)}^k{\bar N}_{(t)}^s
{\tilde h}_{(t)ks}{\Big ]}=0, \qquad {\Rightarrow} \qquad
{\bar N}_{(t),t}^s=-{\frac{1}{2}}{\bar N}_{(t)}^k
{\tilde h}_{(t)}^{is}{\tilde h}_{(t)ik,t},
\ena
and
\eqa 
{\frac{\partial}{{\partial}t}}{\Big [}N_{(t)}^kN_{(t)}^s
{\hat h}_{(t)ks}{\Big ]}=0, \qquad {\Rightarrow} \qquad
N_{(t),t}^s=-{\frac{1}{2}}N_{(t)}^k{\hat h}_{(t)}^{is}{\hat h}_{(t)ik,t}.
\ena
Given the requirements (4) and (5), the conditions shown in equation (3) 
now yield the in general nonzero extra connection coefficients (using a GTCS)
\eqa
{\topstar{\bar {\Gamma}}}_{t0}^{{\,}0}={\frac{{\bar N}_{t},_t}{{\bar N}_t}},
\quad 
{\topstar{\bar {\Gamma}}}_{tj}^{{\,}i}=
{\Big (}{\frac{1}{t}}+{\frac{{\bar N}_t,_t}{{\bar N}_t}}{\Big )}{\delta}^i_j
+{\frac{1}{2}}{\tilde h}_{(t)}^{is}{\tilde h}_{(t)sj,t},
\quad 
{\topstar{\Gamma}}_{tj}^{{\,}i}={\frac{1}{t}}{\delta}^i_j+
{\frac{1}{2}}{\hat h}_{(t)}^{is}{\hat h}_{(t)sj,t}.
\ena
Note that all connection coefficients are symmetric in the lower indices.
The equations of motion in $({\cal N},{\bf g}_t)$ are given by [1, 2]
\eqa
{\frac{d^2x^{\mu}}{d{\lambda}^2}}+{\Big (}
{\topstar{\Gamma}}_{t{\nu}}^{{\,}{\mu}}{\frac{dt}{d{\lambda}}}+
{\topstar{\Gamma}}_{{\beta}{\nu}}^{{\,}{\mu}}{\frac{dx^{\beta}}{d{\lambda}}}
{\Big )}{\frac{dx^{\nu}}{d{\lambda}}}
={\Big (}{\frac{d{\tau}_t}{d{\lambda}}}{\Big )}^2a_{(t)}^{\mu}.
\ena
Here, $d{\tau}_t$ is the proper time interval as measured along the curve,
${\lambda}$ is some general affine parameter, and ${\bf a}_t$ is the 
4-acceleration measured along the curve.
\subsection{The extended field equations}
One important postulate of the OQG is that gravitational quantities should be
``formally'' variable when measured in atomic units. This formal variability 
is also a postulate of revised quasi-metric gravity and applies to all 
dimensionful gravitational quantities. Said formal variability may be viewed as
an interpretation of equation (1) and is directly connected to the spatial 
scale factor ${\bar F}_t{\equiv}{\bar N}_tct$ of the FHSs [1, 2]. 
In particular, the formal variability applies to any potential gravitational 
coupling parameter $G_t$. It is convenient to transfer the formal variability 
of $G_t$ to mass (and charge, if any) so that all formal variability is taken 
into account and included in the {\em active stress-energy tensor} ${\bf T}_t$,
which is the object that couples to space-time geometry via field equations. 
However, dimensional analysis yields that the gravitational coupling must be 
non-universal, i.e., that the electromagnetic active stress-energy tensor 
${\bf T}_t^{{\rm (EM)}}$ and the active stress-energy tensor for material 
particles ${\bf T}_t^{\rm (MA)}$ couple to space-time curvature via two different 
(constant) coupling parameters $G^{\rm B}$ and $G^{\rm S}$, respectively. This 
non-universality of the gravitational coupling is required for consistency 
reasons. As a consequence, compared to GR, the non-universal gravitational 
coupling yields a modification of the right hand side of any quasi-metric 
gravitational field equations. (Said modification was missed in the original 
formulation of quasi-metric gravity.) The quantities $G^{\rm B}$ and $G^{\rm S}$
play the roles of gravitational constants measured in some local gravitational
measurements at some chosen event at the arbitrary reference epoch $t_0$.

Before trying to construct quasi-metric field equations, we notice that we
cannot use curvature tensors calculated from the full connection in 
$({\cal N},{\bf {\bar g}}_t)$ since its dependence on $t$ should not have 
anything directly to do with gravitation. Rather, we must use curvature tensors
calculated from the usual Levi-Civita connection in 
$({\cal M},{\bf {\bar g}}_t)$, i.e., such tensors should be calculated from 
equation (1) holding $t$ fixed. When $t$ varies, said curvature tensors 
constitute tensor families in $({\cal N},{\bf {\bar g}}_t)$. Potential field 
equations in $({\cal N},{\bf {\bar g}}_t)$ may then be found by using 
projections of said curvature tensor families with respect to the FHSs and 
coupling said projections to the relevant projections of ${\bf T}_t^{\rm (EM)}$ 
and ${\bf T}_t^{\rm (MA)}$.

As mentioned earlier, the form of equation (1), and thus of ${\bf {\bar g}}_t$, 
in the OQG was too restricted to admit the existence of a full coupling 
between space-time curvature and the active stress-energy tensor ${\bf T}_t$. 
Rather, a subset of the projected Einstein field equations (with the right hand
sides modified) was tailored to ${\bf {\bar g}}_t$, yielding a {\em partial} 
coupling to space-time curvature of ${\bf T}_t^{{\rm (EM)}}$ and 
${\bf T}_t^{\rm (MA)}$. That is, a postulate of the OQG was the field equation
\eqa
2{\bar R}_{(t){\bar {\perp}}{\bar {\perp}}}=
2(c^{-2}{\bar a}_{{\cal F}{\mid}s}^s+
c^{-4}{\bar a}_{{\cal F}s}{\bar a}_{\cal F}^s-
{\bar K}_{(t)ks}{\bar K}_{(t)}^{ks}+
{\cal L}_{{\bf {\bar n}}_t}{\bar K}_t) \nonumber \\
={\kappa}^{\rm B}(T^{{\rm (EM)}}_{(t){\bar {\perp}}{\bar {\perp}}}
+{\hat T}^{{\rm (EM)}s}_{(t)s})+
{\kappa}^{\rm S}(T^{{\rm (MA)}}_{(t){\bar {\perp}}{\bar {\perp}}}
+{\hat T}^{{\rm (MA)}s}_{(t)s}), \qquad c^{-2}{\bar a}_{{\cal F}s}{\equiv}
{\frac{{\bar N}_t,_s}{{\bar N}_t}}.
\ena
Here, ${\bf {\bar R}}_t$ is the Ricci tensor family corresponding to the metric
family ${\bf {\bar g}}_t$ and the symbol '${\bar {\perp}}$' denotes a scalar
product with $-{\bf {\bar n}}_t$. Moreover, ${\cal L}_{{\bf {\bar n}}_t}$ denotes 
a projected Lie derivative in the direction normal to the FHSs, 
${\bf {\bar K}}_t$ denotes the extrinsic curvature tensor family (with trace 
${\bar K}_t$) of the FHSs, a ``hat'' denotes an object projected into the FHSs 
and the symbol '${\mid}$' denotes a space covariant derivative. (Note that 
${\cal L}_{{\bf {\bar n}}_t}$ operates on space objects only.) Finally
${\kappa}^{\rm B}{\equiv}8{\pi}G^{\rm B}/c^4$ and
${\kappa}^{\rm S}{\equiv}8{\pi}G^{\rm S}/c^4$, where the values of $G^{\rm B}$ and 
$G^{\rm S}$ are by convention chosen as those measured in some local 
gravitational measurements at some chosen event at the arbitrary reference 
epoch $t_0$. 

Except for the non-universal coupling, the field equation (8) is similar to 
its counterpart among the various projections of the Einstein field equations 
in canonical GR. Now it would seem natural to postulate a second set of field 
equations, also yielding a natural correspondence with GR, by adopting those
projections of the Einstein equations involving the quantity
${\bar R}_{(t)j{\bar {\perp}}}$. That is, it would be tempting to postulate a 
coupling of ${\bar R}_{(t)j{\bar {\perp}}}$ directly to 
${\bar T}^{{\rm (EM)}}_{(t)j{\bar {\perp}}}$ and 
${\bar T}^{{\rm (MA)}}_{(t)j{\bar {\perp}}}$. This was indeed done in the OQG. 
However, it was missed that this approach unfortunately does not work since it 
can be shown that this choice implies that a subset of the local conservation
laws and the corresponding subset of the contracted Bianchi identities (i.e., 
equations (26) and (15) below), would be inconsistent in the weak-field limit.

To arrive at somewhat similar field equations but such that no obvious 
inconsistencies appear, an alternative approach will work. First we define 
the vector field family ${\bf {\bar m}}_t$ by its components expressed in a 
GTCS, i.e.,
\eqa
{\bf {\bar m}}_t{\equiv}
-{\frac{1}{{\bar N}_t}}{\frac{\partial}{{\partial}x^0}}-{\frac{t_0}{t}}
{\frac{{\bar N}^i_{(t)}}{{\bar N}_t}}{\frac{\partial}{{\partial}x^i}}=
-{\bf {\bar n}}_t-2{\frac{t_0}{t}}{\frac{{\bar N}^i_{(t)}}{{\bar N}_t}}
{\frac{\partial}{{\partial}x^i}}, \nonumber \\ 
{\bar m}_{(t)}^{\nu}{\bar m}_{(t){\nu}}=
-1+4{\bar N}^i_{(t)}{\bar N}^k_{(t)}{\tilde h}_{(t)ik}, \qquad
{\bar m}_{(t)}^{\nu}{\bar n}_{(t){\nu}}=1.
\ena
Next we use equation (9) to define the space tensor family ${\bf {\bar L}}_t$
via its components in a GTCS (where ${\bar h}_{(t)ij}$ are the components of the
metric family ${\bf {\bar h}}_t$ intrinsic to the FHSs), i.e.,
\eqa
{\bar L}_{(t)ij}{\equiv}
-{\frac{1}{2{\bar N}_t}}{\cal L}_{{\bar N}_t{\bf {\bar m}}_t}
{\bar h}_{(t)ij}={\bar K}_{(t)ij}+
{\frac{1}{{\bar N}_t}}{\frac{\partial}{{\partial}x^0}}{\bar h}_{(t)ij}, \quad
{\bar L}_t{\equiv}{\bar K}_t+{\frac{{\bar h}_{(t)}^{ik}}{{\bar N}_t}}
{\frac{\partial}{{\partial}x^0}}{\bar h}_{(t)ik}.
\ena
One may interpret ${\bf {\bar L}}_t$ as some sort of ``time-reversed''
extrinsic curvature tensor family. The wanted field equation set, having
the properties mentioned above, is then obtained by coupling matter fields to
the quantity ${\bar L}_{(t)j{\mid}s}^s-{\bar L}_t,_j$ rather than to
${\bar R}_{(t)j{\bar {\perp}}}={\bar K}_{(t)j{\mid}s}^s-{\bar K}_t,_j$, i.e.,
\eqa
{\bar R}_{(t)j{\bar {\perp}}}+{\Big (}{\frac{{\bar h}_{(t)}^{ik}}{{\bar N}_t}}
{\frac{\partial}{{\partial}x^0}}{\bar h}_{(t)ij}{\Big )}_{{\mid}k}-
{\Big (}{\frac{{\bar h}_{(t)}^{ik}}{{\bar N}_t}}
{\frac{\partial}{{\partial}x^0}}{\bar h}_{(t)ik}{\Big )},_j
={\bar L}_{(t)j{\mid}i}^i-{\bar L}_t,_j 
={\kappa}^{\rm B}T^{{\rm (EM)}}_{(t)j{\bar {\perp}}}
+{\kappa}^{\rm S}T^{\rm (MA)}_{(t)j{\bar {\perp}}}.
\ena
Equations (8) and (11) consist of one dynamical scalar equation and one 
constraint 3-vector equation, respectively. The dynamical fields in 
$({\cal N},{\bar {\bf g}}_t)$ are the lapse function family ${\bar N}_t$
and the space metric family ${\bf {\tilde h}}_t$. That is, the time evolution 
of ${\bf {\bar K}}_t$ (with $t$ fixed) is determined by the time evolution of 
${\bf {\bar h}}_t$ (with $t$ fixed), since we have that 
$2{\bar K}_{(t)ij}=-{\frac{1}{{\bar N}_t}}{\cal L}_{{\bar N}_t{\bf {\bar n}}_t}
{\bar h}_{(t)ij}$ (see, e.g., [4]). (In addition, the matter variables evolve in
time according to the local conservation laws in $({\cal N},{\bar {\bf g}}_t)$,
see equations (26) and (27) below.) Unfortunately, the equation set (8), (11) 
has no (scalar) wave-like solutions in vacuum, given the restricions on 
${\bf {\bar g}}_t$ from the OQG. Thus using the approach of the OQG, no aspects
of ${\bf {\bar g}}_t$ were left free, meaning that ${\bf {\bar g}}_t$ would be 
fully determinable by the matter sources alone. So the OQG is essentially a 
waveless approximation theory, and it must therefore be discarded as a 
potentially viable candidate for quasi-metric gravity.

To correct said inadequacies of the OQG, it is crucial to find new field 
equations that allow the existence of GR-like gravitational waves. Since it is 
necessary to have a correspondence between the new field equations and the OQG,
it is necessary to keep equations (8) and (11) and to extend them with 
additional field equations. One might expect that an extended set should 
represent a full coupling between space-time curvature and ${\bf T}_t$, in 
addition to being compatible with equation (1). That is, we would expect to 
find a new space-time tensor family ${\bf {\bar Q}}_t$ defined from its 
projections ${\bar Q}_{(t){\bar {\perp}}{\bar {\perp}}}$,
${\bar Q}_{(t)j{\bar {\perp}}}={\bar Q}_{(t){\bar {\perp}}j}$ and ${\bar Q}_{(t)ij}$ 
with respect to the FHSs. These projections are expected to play almost the 
same role as do the projections of the Einstein tensor in canonical GR. 
However, it is important to notice that unlike ${\bf {\bar R}}_t$ and the 
Einstein tensor family ${\bf {\bar G}}_t$, any definition of ${\bf {\bar Q}}_t$
will depend directly on the geometry of the FHSs and their extrinsic curvature.
This means that the expected expressions for the projections of 
${\bf {\bar Q}}_t$ will not be exactly valid for any hypersurfaces other than 
the FHSs. In contrast, in canonical GR, the projections of the Einstein tensor 
${\bf {\bar G}}$ on a Lorentzian manifold with metric ${\bf {\bar g}}$ is 
valid for any foliation of ${\bf {\bar g}}$ into spatial hypersurfaces. The 
quasi-metric counterpart to these projections is the projections of of 
${\bf {\bar G}}_t$ with respect to the FHSs and they take this common general 
form (see e.g., [2] and references therein)
\eqa
{\bar G}_{(t){\bar {\perp}}{\bar {\perp}}}={\frac{1}{2}}({\bar P}_t
+{\bar K}_t^2-{\bar K}_{(t)ks}{\bar K}_{(t)}^{ks}), 
\ena
\eqa
{\bar G}_{(t){\bar {\perp}}j}{\equiv}{\bar R}_{(t){\bar {\perp}}j}
=({\bar K}_{(t)j}^k-{\bar K}_t{\delta}_{{\ }j}^k)_{{\mid}k},
\ena
\eqa
{\bar G}_{(t)ij}=-{\frac{1}{{\bar N}_t}}
{\cal L}_{{\bar N}_t{\bf {\bar n}}_t}({\bar K}_{(t)ij}-
{\bar K}_t{\bar h}_{(t)ij})+3{\bar K}_t{\bar K}_{(t)ij}-
{\frac{1}{2}}({\bar K}_t^2+
{\bar K}_{(t)ks}{\bar K}_{(t)}^{ks}){\bar h}_{(t)ij} \nonumber \\
-2{\bar K}_{(t)is}{\bar K}_{(t)j}^s
-c^{-2}{\bar a}_{{\cal F}i{\mid}j}-c^{-4}{\bar a}_{{\cal F}i}
{\bar a}_{{\cal F}j}+(c^{-2}{\bar a}^s_{{\cal F}{\mid}s}
+c^{-4}{\bar a}_{\cal F}^s{\bar a}_{{\cal F}s}){\bar h}_{(t)ij}+
{\bar H}_{(t)ij}.
\ena
Here, ${\bar P}_t$ and ${\bar H}_{(t)ij}$ are the Ricci scalar family and the 
components of the Einstein tensor family ${\bf {\bar H}}_t$ intrinsic to to 
the FHSs, respectively.

We will now require that ${\bf {\bar Q}}_t$ and ${\bf {\bar G}}_t$ should have 
somewhat similar dynamical structures. That is, ${\bar Q}_{(t)ij}$ and 
${\bar G}_{(t)ij}$ should both predict weak GR-like gravitational waves in 
vacuum via having common (up to signs) second-order terms $-{\frac{1}{\bar N}}
{\cal L}_{{\bar N}{\bf {\bar n}}}{\bar K}_{(t)ij}$ and ${\bar H}_{(t)ij}$ in equation 
(14). Furthermore, we must have that ${\bar Q}_{(t){\bar {\perp}}{\bar {\perp}}}+
{\hat{\bar Q}}^s_{(t)s}=2{\bar R}_{(t){\bar {\perp}}{\bar {\perp}}}$ to fulfil equation
(8), and ${\bar Q}_{(t){\bar {\perp}}j}={\bar L}_{(t)j{\mid}i}^i-{\bar L}_t,_j$ to 
fulfil equation (11). Besides, the extended field equations should also yield 
the same solutions as the OQG for the metrically static vacuum cases (for which
the extrinsic curvature vanishes identically). Thus for these cases, the 
equation ${\bar Q}_{(t)ij}=0$ should yield the relationship ${\bar H}_{(t)ij}+
c^{-2}{\bar a}_{{\cal F}i{\mid}j}+c^{-4}{\bar a}_{{\cal F}i}{\bar a}_{{\cal F}j}-
(c^{-2}{\bar a}_{{\cal F}{\mid}s}^s-{\frac{1}{(ct{\bar N}_t)^2}}){\bar h}_{(t)ij}=0$,
which follows directly from the OQG [2]. But the extrinsic curvature also 
vanishes identically for metrically static interiors, so this means that we 
should have ${\bar Q}_{(t)ij}=-c^{-2}{\bar a}_{{\cal F}i{\mid}j}-
c^{-4}{\bar a}_{{\cal F}i}{\bar a}_{{\cal F}j}+(c^{-2}{\bar a}_{{\cal F}{\mid}s}^s
-{\frac{1}{(ct{\bar N}_t)^2}}){\bar h}_{(t)ij}-{\bar H}_{(t)ij}$ and thus
${\bar Q}_{(t){\bar {\perp}}{\bar {\perp}}}=-{\frac{1}{2}}{\bar P}_t
+3c^{-4}{\bar a}_{{\cal F}s}{\bar a}_{\cal F}^s+{\frac{3}{(ct{\bar N}_t)^2}}$
for the metrically static cases. (The other sign for ${\bar Q}_{(t)ij}$ cannot 
be chosen since we for physical reasons in general must have that
${\bar Q}_{(t){\bar {\perp}}{\bar {\perp}}}>0$ and ${\hat{\bar Q}}^s_{(t)s}>0$ for 
metrically static interiors. That is, we expect these quantities to be 
non-negative since they should be coupled to suitable projections of 
${\bf T}_t$.)

However, at this point a crucial problem arises due to the contracted Bianchi
identities ${\bar G}_{(t){\mu};{\nu}}^{\nu}{\equiv}0$ (where a semicolon denotes 
taking a metric covariant derivative in component notation, with $t$ fixed). 
Projected with respect to the FHSs, these identities read (see, e.g., [4])
\eqa
{\cal L}_{{\bf {\bar n}}_t}{\bar G}_{(t){\bar {\perp}}{\bar {\perp}}}=
{\bar K}_t{\bar G}_{(t){\bar {\perp}}{\bar {\perp}}}+{\bar K}_{(t)}^{ks}
{\bar G}_{(t)ks}-2c^{-2}{\bar a}_{\cal F}^s{\bar G}_{(t){\bar {\perp}}s}
-{\hat {\bar G}}^s_{(t){\bar {\perp}}{\mid}s},
\ena
\eqa
{\frac{1}{{\bar N}_t}}{\cal L}_{{\bar N}_t{\bf {\bar n}}_t}
{\bar G}_{(t)j{\bar {\perp}}}={\bar K}_t{\bar G}_{(t)j{\bar {\perp}}}
-c^{-2}{\bar a}_{{\cal F}j}{\bar G}_{(t){\bar {\perp}}{\bar {\perp}}}
-c^{-2}{\bar a}_{\cal F}^s{\bar G}_{(t)sj}-{\hat {\bar G}}^s_{(t)j{\mid}s}.
\ena
That is, it turns out that equations (11) and (16), in combination with the 
deduced expressions for ${\bar Q}_{(t){\bar {\perp}}{\bar {\perp}}}$ and 
${\bar Q}_{(t)ij}$ for metrically static interiors, yield the wrong Newtonian 
limit, so that equation (16) does not correspond with its counterpart Euler 
equation (see section 2.3). In fact, the only way to avoid said problem while 
still keeping the relationship ${\bar Q}_{(t){\bar {\perp}}{\bar {\perp}}}+
{\hat{\bar Q}}^s_{(t)s}=2{\bar R}_{(t){\bar {\perp}}{\bar {\perp}}}$ is to set 
${\bar Q}_{(t){\bar {\perp}}{\bar {\perp}}}=2{\bar R}_{(t){\bar {\perp}}{\bar {\perp}}}$ and 
${\bar Q}_{(t)ks}{\bar h}_{(t)}^{ks}=0$, with the extra 
condition ${\hat {\bar Q}}^s_{(t)j{\mid}s}-c^{-2}{\bar a}_{\cal F}^s{\bar Q}_{(t)js}
=0$ coming from equation (16). However, this yields no possible consistent 
coupling of $T_{(t)ij}$ to ${\bar Q}_{(t)ij}$ given equation (27) below, so to 
avoid said problem we are forced to set ${\bar Q}_{(t)ij}=0$. Thus there can be 
no extra scalar field equation besides equation (8) and also no additional 
spatial tensor equation representing a full coupling to the spatial projections
of ${\bf T}_t$. In other words, we have found that {\em it is not possible to 
construct a viable, fully coupled quasi-metric gravitational theory.} 

Nevertheless, fortunately it is still possible to have a partially coupled,
manifestly traceless field equation ${\bar Q}_{(t)ij}=0$ for the general case. 
Such a field equation will have the desired dynamical properties in addition 
to being compatible with equation (16) (the couplig to ${\bf T}_t$ is via 
equation (8)). The choice of terms quadratic in extrinsic curvature in such an 
equation would seem somewhat uncertain, but this question can be resolved by
a restriction involving a particular projection of the Weyl tensor family 
${\bf {\bar C}}_t$. That is, we require that the projection
${\bar C}_{(t){\bar {\perp}}i{\bar {\perp}}j}$ should be determined from the intrinsic
geometry of the FHSs alone, with no explicit dependence on extrinsic curvature
(or on ${\bf {\bar a}}_{\cal F}$). Thus we define a (unique) relationship having
this property, i.e.,
\eqa
{\bar C}_{(t){\bar {\perp}}i{\bar {\perp}}j}={\tilde H}_{(t)ij}+
{\frac{1}{(ct{\bar N}_t)^2}}{\bar h}_{(t)ij},
\ena
where ${\bf {\tilde H}}_t$ is the spatial Einstein tensor family calculated 
from the metric family ${\bf {\tilde h}}_t$. (Note that the 
foliation-dependence of equation (17) (and thus of the field equations) is 
directly given from its right hand side.) Moreover, we also have in general 
that ${\bf {\bar C}}_t$ can be expressed by the Riemann tensor family, the 
Ricci tensor family and the Ricci scalar family ${\bar R}_t$. In particular, 
this yields (see, e.g., [4])
\eqa
{\bar C}_{(t){\bar {\perp}}i{\bar {\perp}}j}=
{\bar R}_{(t){\bar {\perp}}i{\bar {\perp}}j}+{\frac{1}{2}}{\bar R}_{(t)ij}
-{\frac{1}{2}}{\Big (}{\bar R}_{(t){\bar {\perp}}{\bar {\perp}}}+{\frac{1}{3}}
{\bar R}_t{\Big )}{\bar h}_{(t)ij}, \nonumber \\
{\bar R}_{(t){\bar {\perp}}i{\bar {\perp}}j}={\frac{1}{{\bar N}_t}}
{\cal L}_{{\bar N}_t{\bf {\bar n}}_t}{\bar K}_{(t)ij}+
{\bar K}_{(t)i}^{s}{\bar K}_{(t)sj}+c^{-2}{\bar a}_{{\cal F}i{\mid}j}+
c^{-4}{\bar a}_{{\cal F}i}{\bar a}_{{\cal F}j}, \nonumber \\
{\bar R}_t={\bar P}_t-2{\cal L}_{{\bf {\bar n}}_t}{\bar K}_t
+{\bar K}_{(t)ks}{\bar K}_{(t)}^{ks}+{\bar K}_t^2
-2c^{-2}{\bar a}_{{\cal F}{\mid}s}^s
-2c^{-4}{\bar a}_{{\cal F}}^s{\bar a}_{{\cal F}s}.
\ena
Equations (18) may now be inserted into equation (17) to give a definition of 
${\bar Q}_{(t)ij}$ via the quantities ${\bar G}_{(t)ij}$, 
${\bar R}_{(t){\bar {\perp}}{\bar {\perp}}}$ and 
${\bar G}_{(t){\bar {\perp}}{\bar {\perp}}}$. That is, we define ${\bar Q}_{(t)ij}$ 
from equation (14) and the requirement that
\eqa
{\bar G}_{(t)ij}=-{\bar Q}_{(t)ij}-2c^{-2}{\bar a}_{{\cal F}i{\mid}j}-
2c^{-4}{\bar a}_{{\cal F}i}{\bar a}_{{\cal F}j}
-2{\bar K}_{(t)i}^{s}{\bar K}_{(t)sj}+2{\bar K}_t{\bar K}_{(t)ij}
\nonumber \\
+{\frac{1}{3}}{\Big [}2{\bar R}_{(t){\bar {\perp}}{\bar {\perp}}}
-{\bar G}_{(t){\bar {\perp}}{\bar {\perp}}}+2c^{-2}{\bar a}_{{\cal F}{\mid}s}^s
+2c^{-4}{\bar a}_{{\cal F}}^s{\bar a}_{{\cal F}s}
+2{\bar K}_{(t)ks}{\bar K}_{(t)}^{ks}-2{\bar K}_t^2
{\Big ]}{\bar h}_{(t)ij}.
\ena
We then get the definition (note the prior-geometric term)
\eqa
{\bar Q}_{(t)ij}{\equiv}{\frac{1}{{\bar N}_t}}{\cal L}_{{\bar N}_t
{\bf {\bar n}}_t}{\bar K}_{(t)ij}+
{\frac{1}{3}}{\Big [}2{\bar K}_{(t)ks}{\bar K}_{(t)}^{ks}-{\bar K}_t^2
-{\cal L}_{{\bf {\bar n}}_t}{\bar K}_t
{\Big ]}{\bar h}_{(t)ij}
\nonumber \\
+{\bar K}_t{\bar K}_{(t)ij}-c^{-2}{\bar a}_{{\cal F}i{\mid}j}-
c^{-4}{\bar a}_{{\cal F}i}{\bar a}_{{\cal F}j}
+{\Big [}c^{-2}{\bar a}_{{\cal F}{\mid}s}^s
-{\frac{1}{(ct{\bar N}_t)^2}}{\Big ]}{\bar h}_{(t)ij}-{\bar H}_{(t)ij}=0,
\ena
where the requirement on the spatial Ricci curvature scalar family 
${\bar P}_t$,
\eqa
{\bar P}_t=-4c^{-2}{\bar a}_{{\cal F}{\mid}s}^s
+2c^{-4}{\bar a}_{{\cal F}}^s{\bar a}_{{\cal F}s}+{\frac{6}{(ct{\bar N}_t)^2}},
\ena
ensures that equation (20) is indeed manifestly traceless. Besides, the 
components of the spatial Einstein tensor family ${\bf {\bar H}}_t$ are given 
by
\eqa
{\bar H}_{(t)ij}=-c^{-2}{\bar a}_{{\cal F}i{\mid}j}-
c^{-4}{\bar a}_{{\cal F}i}{\bar a}_{{\cal F}j}+
c^{-2}{\bar a}_{{\cal F}{\mid}s}^s{\bar h}_{(t)ij}+{\tilde H}_{(t)ij}.
\ena 
Note that, while equation (21) implies that ${\tilde P}_t={\frac{6}{(ct_0)^2}}$
is fixed by the prior geometry, ${\tilde H}_{(t)ij}$ is not necessarily equal to
the prior-geometric quantity $-{\frac{1}{(ct_0)^2}}{\tilde h}_{(t)ij}$. This 
shows that, while there is prior 3-geometry, there is still some dynamical 
freedom associated with the metric family ${\bf {\tilde h}}_t$. This is further 
illustrated by writing equation (20) in the form (using equations (8) and (22))
\eqa
{\frac{1}{{\bar N}_t}}{\cal L}_{{\bar N}_t{\bf {\bar n}}_t}{\bar K}_{(t)ij}
+{\bar K}_t{\bar K}_{(t)ij}-{\tilde H}_{(t)ij} \nonumber \\
={\frac{1}{3}}{\Big [}{\bar R}_{(t){\bar {\perp}}{\bar {\perp}}}
+{\bar K}_t^2-{\bar K}_{(t)ks}{\bar K}_{(t)}^{ks}
-c^{-2}{\bar a}_{{\cal F}{\mid}s}^s
-c^{-4}{\bar a}_{{\cal F}}^s{\bar a}_{{\cal F}s}+{\frac{3}{(ct{\bar N}_t)^2}}
{\Big ]}{\bar h}_{(t)ij}.
\ena
We notice that taking the trace of equation (23) recovers the (general)
expression (8) for ${\bar R}_{(t){\bar {\perp}}{\bar {\perp}}}$.
Equations (8), (11) and (20) determine ${\bar Q}_{(t){\bar {\perp}}{\bar {\perp}}}
{\equiv}2{\bar R}_{(t){\bar {\perp}}{\bar {\perp}}}$, ${\bar Q}_{(t){\bar {\perp}}j}
{\equiv}{\bar L}_{(t)j{\mid}i}^i-{\bar L}_t,_j$ and ${\bar Q}_{(t)ij}$, 
respectively. This yields 9 restricions on the the 20 independent components of
the Riemann tensor family in 4 dimensions. Equation (21) yields one extra 
restriction so that all together, said equations yield 10 restricions on said 
components, the same number as for the full Einstein tensor in ordinary GR.
Besides, for quasi-metric gravity we see from equation (17) that we get 5 
restrictions on the 10 independent components of the Weyl tensor family in 
addition to 5 restrictions on the 10 independent components of the Ricci tensor
family. On the other hand, in GR the field equations determine the Ricci 
tensor in full, leaving the Weyl tensor free.

The full set of quasi-metric field equations then consists of equations (8),
(11), (20) and (21). (Equation (17) or (23) may alternatively be substituted 
for equation (20).) Note that these quasi-metric field equations have a 
somewhat similar split-up as Einstein's field equations into dynamical 
equations and constraints. That is, equations (11) and (21) represent 4 
constraint equations while equations (8) and (18) represent 6 dynamical 
equations, the same numbers as for GR. However, the Einstein equations include 
no counterpart to equation (21) but rather an extra scalar constraint 
corresponding to equation (12); such an equation is missing in quasi-metric 
gravity. Besides, equation (20) (or (23)) is only partially coupled to matter 
sources. In this context it is useful to compare equations (14) and (23); while
the former is fully coupled to the spatial projection $T_{(t)ij}$ of the 
stress-energy tensor via the Einstein field equations, the latter is obviously 
not fully coupled to matter sources since such a coupling enters equation (23) 
only via the scalar quantity ${\bar R}_{(t){\bar {\perp}}{\bar {\perp}}}$. This 
property of equation (23) means that interior solutions will be less dependent 
on the source's equation of state than for comparable situations in GR, so that
any quasi-metric interior solution should cover a wider range of physical 
conditions than its counterparts in GR.

A useful coordinate expression for ${\bf {\bar K}}_t$ is the well-known 
(except for the $t$-dependence) formula from canonical GR
\eqa
{\bar K}_{(t)ij}={\frac{1}{2{\bar N}_t}}{\Big [}{\frac{t}{t_0}}
({\bar N}_{(t)i{\mid}j}+{\bar N}_{(t)j{\mid}i})-
{\frac{\partial}{{\partial}x^0}}{\bar h}_{(t)ij}{\Big ]}.
\ena
Note that ${\bar K}_{(t)ij}$ contains first-order but not second-order time 
derivatives of ${\bar h}_{(t)ij}$. Moreover, constraint equations on an initial
FHS are determined by the initial data and do by definition not contain 
second-order time derivatives. On the other hand, this is the role of the 
dynamical equations via terms like ${\frac{1}{{\bar N}_t}}{\cal L}
_{{\bar N}_t{\bf {\bar n}}_t}{\bar K}_{(t)ij}$ or ${\cal L}_{{\bf {\bar n}}_t}{\bar K}_t$. 
Now we see that the quantities ${\bar G}_{(t){\bar {\perp}}{\bar {\perp}}}$ and 
${\bar L}_{(t)j{\mid}i}^i-{\bar L}_t,_j$ are both determined by the initial data, 
while the quantities ${\bar R}_{(t){\bar {\perp}}{\bar {\perp}}}$, ${\bar G}_{(t)ij}$ 
and ${\bar Q}_{(t)ij}$ are not. 
 
Next we have the original local conservation laws of ${\bf T}_t$ in 
$({\cal N},{\bf {\bar g}}_t)$. These are unchanged from the OQG, i.e., for 
fixed $t$ we have
\eqa
T_{(t){\mu};{\nu}}^{\nu}=2{\frac{{\bar N}_t,_{\nu}}{{\bar N}_t}}
T_{(t){\mu}}^{\nu}=2c^{-2}{\bar a}_{{\cal F}s}{\hat T}_{(t){\mu}}^s
-2{\frac{{\bar N}_t,_{\bar {\perp}}}{{\bar N}_t}}T_{(t){\bar {\perp}}{\mu}}.
\ena
By projecting equation (25) with respect to the FHSs we get (see, e.g., [4] for
general projection formulae)
\eqa
{\cal L}_{{\bf {\bar n}}_t}T_{(t){\bar {\perp}}{\bar {\perp}}}=
{\Big (}{\bar K}_t-2{\frac{{\bar N}_t,_{\bar {\perp}}}{{\bar N}_t}}
{\Big )}T_{(t){\bar {\perp}}{\bar {\perp}}}
+{\bar K}_{(t)ks}{\hat T}_{(t)}^{ks}-{\hat T}^s_{(t){\bar {\perp}}{\mid}s},
\ena
\eqa
{\frac{1}{{\bar N}_t}}{\cal L}_{{\bar N}_t{\bf {\bar n}}_t}
T_{(t)j{\bar {\perp}}}={\Big (}{\bar K}_t
-2{\frac{{\bar N}_t,_{\bar {\perp}}}{{\bar N}_t}}
{\Big )}T_{(t)j{\bar {\perp}}}-c^{-2}{\bar a}_{{\cal F}j}
T_{(t){\bar {\perp}}{\bar {\perp}}}
+c^{-2}{\bar a}_{{\cal F}s}{\hat T}_{(t)j}^s-{\hat T}^s_{(t)j{\mid}s}.
\ena
The conservation laws must take the form (25) to be consistent with classical 
electrodynamics coupled to quasi-metric gravity [5]. However, equation (25) and
equations (26)-(27) apply to both ${\bf T}_t^{{\rm (EM)}}$ and 
${\bf T}_t^{\rm (MA)}$ alike. Moreover, if the only $t$-dependence of ${\bf T}_t$
is via the above-mentioned formal variability, ${\bf T}_t$ is locally conserved
when $t$ varies as well. Note that, the covariantly conserved quantity 
following from equation (25) is ${\bar N}_t^{-2}{\bf T}_t$ rather than 
${\bf T}_t$. But since ${\bar N}_t^{-2}{\bf T}_t$ depends on the distinguished 
foliation of quasi-metric space-time into the FHSs, this means that potential 
field equations cannot be found from an invariant action principle obtained 
from any Lagrangian involving only ${\bf {\bar g}}_t$ and its derivatives, with
no dependence on any particular foliation. We thus have, unlike its counterpart
in GR, that equation (25) does not automatially follow from the field 
equations. That is, equation (25) represents real restrictions on what kind of 
sources can be admitted in the field equations for a given gravitational 
system. 

Next we notice that, when specifying initial data ${\bf {\bar h}}_t$,
${\bf {\bar L}}_t$ and ${\bf {\bar K}}_t$ on an initial FHS, due to equation 
(21) there is no freedom to choose the lapse function family ${\bar N}_t$ 
independently. However, since the unit normal vector field family 
${\bf {\bar n}}_t$ uniquely determines the world lines of the FOs, there is no 
freedom to choose the components ${\frac{t_0}{t}}{\bar N}_{(t)}^j$ of the shift 
vector family independently either. But this means, unlike the GR case, that 
the quasi-metric initial-value system describes the time evolution of a fixed 
sequence of spatial hypersurfaces, i.e., the FHSs. That is, there is no gauge 
freedom to choose lapse and shift as for the GR case, where the evolution of an
initial spatial hypersurface into some fixed final one may be done by foliating 
space-time in many different ways. This means that equation (25) is required
to hold, independently of the field equations, at every subsequent FHS of the
quasi-metric initial-value problem. Thus, the dynamical equations cannot 
automatically preserve the constraints since equation (21) eliminates any gauge
freedom in choosing lapse and shift. On the other hand, it is well known that 
the vacuum Einstein field equations preserve the constraints in GR since the 
Bianchi identities assure that there are no extra restrictions.

Finally we notice that the quantities ${\bar N}_t,_t$ and ${\tilde h}_{(t)ij},_t$ 
play no dynamical role in the quasi-metric initial-value problem since in 
principle, they can be chosen freely on an initial FHS, yet their values at 
subsequent FHSs cannot be determined from dynamical equations. Rather, to 
control the evolution of ${\bar N}_t$ and ${\tilde h}_{(t)ij}$, the values of 
said quantities must be determined independently from indirect effects of the 
cosmic expansion on the matter source for each time step. An example of this is 
given in section 3.1 below.

The field equations make it possible to calculate ${\bf {\bar g}}_t$ from the 
projections of the physical source ${\bf T}_t$ with respect to the FHSs. 
Moreover, said equations are in principle valid only for the FHSs as 
long as the global time function $t$ is unique. However, the uniqueness of $t$ 
follows from the topological structure of quasi-metric space-time, since the 
FHSs are defined to be compact with positive curvature scalar ${\tilde P}_t$.  
In quasi-metric cosmology, this singles out the cosmic rest frame (the frame 
where the cosmic relic microwave radiation is measured to be isotropic on 
average) as a natural ``preferred frame'' since the FOs should be at rest on 
average with respect to this frame. Thus when doing cosmology, the global time 
function and the FHSs are given {\em a priori} from the postulated form of 
quasi-metric space-time.

But for local, isolated systems, applications of the field equations would seem
to be limited in practice since they are expressed in terms of one particular 
foliation of quasi-metric space-time into spatial hypersurfaces, apparently 
involving the cosmic rest frame. Therefore, for isolated systems, one may
substitute the condition ${\tilde P}_t={\frac{6}{(ct_0)^2}}$ with the 
approximate alternative condition ${\tilde P}_t=0$. This means that the FHSs 
may be taken to be approximately flat sufficienly far from an isolated system.
But if the FHSs are taken to be asymptotically flat, this means that the global
time function will no longer be unique. In fact, it will then be possible to 
define an alternative global time function $t'={x^0}^{'}/c$ and an alternative 
foliation of ${\bf {\bar g}}_{t^{'}}$ into an alternative set of spatial 
hypersurfaces (also being asymptotically flat). An alternative class of 
observers always moving orthogonally to the alternative hypersurfaces may then 
be defined such that said observers are at rest with respect to the barycentre 
of the isolated system. Moreover, the field equations (with 
${\tilde P}_{t'}=0$) may then be transformed with respect to this new set of 
hypersurfaces. However, the field equations would not be invariant under said 
transformation; they would depend on the velocity of the isolated system 
with respect to the cosmic rest frame. In practice the 
``preferred frame''-effects introduced by said procedure should be small (at 
most of post-Newtonian order), if the size of the isolated system is small 
compared to $ct_0$ and its local speed with respect to the cosmic rest frame is
much smaller than the speed of light.

In this section we have described all necessary changes in the basic equations 
of quasi-metric gravity when switching from the OQG to the revised theory.
There will be no further modifications. In particular, the transformation 
${\bf {\bar g}}_t{\rightarrow}{\bf g}_t$ will be defined as before [1, 2]. In 
this context, we notice that to have the full initial value problem in 
$({\cal N},{\bf g}_t)$, the transformation 
${\bf {\bar g}}_t{\rightarrow}{\bf g}_t$ must be performed at each time step
so that equation (7) can be used to propagate the sources.
\subsection{Weak-field approximations}
It is necessary to clearly define the weak-field approximations of the 
quasi-metric equations as applied to isolated systems. First we approximate
said equations at the Newtonian level of precision. But Newtonian theory should
have a correspondence with the {\em metric} part of quasi-metric theory, and 
not with the non-metric part, which has no Newtonian counterpart. Therefore, a 
more useful approximation than the traditional Newtonian limit can be made by 
taking the weak-field limit of equation (1), but such that the global spatial 
scale factor ${\frac{t}{t_0}}$ is included. That is, in this 
``quasi-Newtonian'' limit, the FHSs are taken as flat, but non-static since the
explicit dependence on $t$ is still present. Besides, in the quasi-Newtonian 
limit, ${\bf {\bar g}}_t={\bf g}_t$ (and we can thus drop the bar labels if 
convenient). 

We may now write down the weak-field limit of equation (1) at the Newtonian 
level of precision. To do that, we estimate the smallness of the terms to be of
the same order as that of the small quantity ${\frac{w}{c}}$ to some power, 
where $w$ is the typical speed of the (gravitating) matter with respect to the 
FOs. For an isolated system, and at the Newtonian level of precision, an 
alternative class of observers being at rest with respect to a suitable GTCS 
using Cartesian coordinates, can be substituted for the FOs (see section 2.2). 
In this GTCS, the quasi-Newtonian metric family then has the components
\eqa
{\bar g}_{(t)00}=-1+2c^{-2}U(x^{\mu})+O(4), \quad
{\bar g}_{(t)i0}={\bar g}_{(t)i0}=0+O(3), {\quad} 
{\bar h}_{(t)ij}={\frac{t^2}{t_0^2}}{\delta}_{ij}+O(2),
\ena
where $-U(x^{\mu})$ is the Newtonian potential. Note that equation (28) is 
consistent with the general metric family (1) since to Newtonian accuracy, 
we can neglect any contribution to ${\bar h}_{(t)ij}$ from $c^{-2}U(x^{\mu})$ 
as this term is of $O(2)$. The quasi-Newtonian form (28) of the metric 
family is useful since it takes sufficiently care of the effects of the global 
cosmic expansion for weak gravitational fields and slow motions. Moreover, the 
traditional Newtonian metric form can be recovered just by setting the factor 
${\frac{t}{t_0}}$ equal to unity in equation (28).

Since the weak-field approximation of the extrinsic curvature tensor family 
${\bf {\bar K}}_t$ is at least of $O(3)$ or higher, it may be neglected in the 
field equations and in the local conservation laws at the Newtonian level of 
precision. Then we see that equation (8) yields Newton's field equation (if
one ignores the contribution 
${\kappa}^{\rm B}(T_{(t){\bar{\perp}}{\bar{\perp}}}^{\rm (EM)}+
{\hat T}^{{\rm (EM)}s}_{(t)s})$ coming from electromagnetic fields) whereas 
equations (11) and (20) become vacuous for sufficiently weak gravitational 
fields. Moreover, it is straightforward to show that for a perfect fluid 
source, to Newtonian accuracy each of equations (26)-(27) corresponds to the 
counterpart Euler equation valid for Newtonian fluid dynamics.

The next level of precision beyond the Newtonian limit is the post-Newtonian
approximation (applied to isolated gravitational systems). At this level of 
precision, a general weak-field approximation formalism valid for the QMF
(somewhat similar to the parametrized post-Newtonian (PPN) formalism valid for 
metric theories of gravity) would be useful for comparing predictions to 
observations. But since such a formalism does not exist, one should have an 
idea of how well a standard PPN-analysis of the quasi-metric field equations 
might work.

Obviously, the PPN-formalism is not designed to take into account the 
non-metric aspects of the QMF. This means that any PPN-analysis of said field 
equations will be limited to their metric approximations. Moreover, as 
discussed in section 2.2, for a sufficiently small isolated system, the global 
curvature of space may be neglected to a good approximation. Then an 
approximately global cosmic frame with an associated approximately global time 
function may be chosen such that the barycentre of the system is taken to be at
rest with respect to this frame (which may be identified with the standard PPN 
coordinate system). The PPN-approximations of the field equations may then be 
transformed to this frame. But they will not be invariant under this 
transformation since said frame represents an alternative foliation of 
space-time into spatial hypersurfaces. That is, in QMR there should be 
``preferred frame''-effects somewhat resembling those covered by the 
PPN-formalism, and with the condition that the PPN-parameter ${\gamma}=-1$ for 
the PPN-metric ${\bf {\bar g}}$. However, one must be careful not to interpret 
said effects as due to a variable gravitational ``constant'' on top of 
Newtonian theory (as is done in a standard PPN-analysis); this would be 
inconsistent with quasi-metric gravity. As a result, the detectability of any 
``preferred frame''-effects should be significantly more subtle for 
quasi-metric gravity than for metric theories of gravity.

However, even metric approximations of the field equations are not very 
suitable for a standard PPN-analysis since the resulting PPN-metric 
${\bf {\bar g}}$ is not the one to which experiments are to be compared, and 
${\bf {\bar g}}$ will {\em not} have an acceptable set of PPN-parameters 
according to metric theory. To have that, the transformation 
${\bf {\bar g}}{\rightarrow}{\bf g}$ must be taken into account. However, 
this transformation has no counterpart in GR (or other metric theories of 
gravity), and the standard PPN-formalism has not been designed to take such a 
feature into account (in fact, said transformation would turn the 
PPN-parameters into scalar fields rather than new constants). Besides, finding 
a complete set of PPN-parameters for ${\bf {\bar g}}$ turns out to be 
problematic since the relationships assumed to hold between said parameters in 
metric theories will not necessarily hold for quasi-metric gravity. This 
typically will lead to inconsistencies. So, as a consequence of all these 
complications, the conclusion is that quasi-metric gravity is unsuitable for a 
full standard PPN-analysis [1].
\section{Two example solutions}
In this section, we find two solutions of the extended field equations for
simple systems. Of these, the cosmological solution has been found previously 
for the OQG and is included here for illustrative purposes. Example solutions 
do not cover metrically static systems since for such systems, the solutions of
the extended field equations and those of the OQG coincide. (This can be seen 
directly from equation (23) since ${\bf {\bar K}}_t$ vanishes identically for 
metrically static systems.) See [5, 6] for some spherically symmetric cases.
\subsection{Isotropic cosmology}
Isotropic cosmology in the OQG has been treated in [3]. Now equation (20) 
yields that the solution found there is the unique solution also of the revised
theory. That is, introducing a spherical GTCS 
${\{ }x^0,{\chi},{\theta},{\phi}{\} }$, for isotropic cosmology equation (21) 
ensures that equation (1) takes the form
\eqa
{\overline {ds}}_t^2={\bar N}_t^2{\Big \{}-(dx^0)^2+(ct)^2
{\Big (}d{\chi}^2+{\sin}^2{\chi}d{\Omega}^2{\Big )}{\Big \}},
\ena
where $d{\Omega}^2{\equiv}d{\theta}^2+{\sin}^2{\theta}d{\phi}^2$. The extrinsic
curvature tensor and the intrinsic curvature of the FHSs obtained from 
equation (29) are given by
\eqa
{\bar K}_{(t)ik}={\frac{{\bar N}_{t,{\bar {\perp}}}}{{\bar N}_t}}
{\bar h}_{(t)ik}, \qquad
{\bar K}_t=3{\frac{{\bar N}_{t,{\bar {\perp}}}}{{\bar N}_t}}, \qquad
{\bar H}_{(t)ik}=-{\frac{1}{(ct{\bar N}_t)^2}}{\bar h}_{(t)ik}, 
\qquad {\bar P}_t={\frac{6}{(ct{\bar N}_t)^2}}.
\ena
Next we assume that the quasi-metric universe is filled with a perfect fluid 
with active mass density ${\tilde {\varrho}}_{\rm m}$ and corresponding pressure
${\tilde p}$, so that
\eqa
T_{(t){\bar {\perp}}{\bar {\perp}}}={\tilde {\varrho}}_{\rm m}c^2{\equiv}
{\Big (}{\frac{t_0}{{\bar N}_tt}}{\Big )}^2{\bar {\varrho}}_{\rm m}(t)c^2,
\qquad T_{(t){\chi}}^{\chi}=T_{(t){\theta}}^{\theta}=T_{(t){\phi}}^{\phi}=
{\tilde p}{\equiv}{\Big (}{\frac{t_0}{{\bar N}_tt}}{\Big )}^2{\bar p}(t),
\ena
where we have set the arbitrary boundary condition ${\bar N}_t(t_0)=1$ for the
present reference epoch $t_0$. Furthermore we have the relationship
\eqa
{\bar {\varrho}}_{\rm m}=
\left\{
\begin{array}{ll}
{\frac{t^3}{t_0^3}}{\bar N}_t^3{\varrho}_{\rm m}
& \text{for a fluid of material particles,} \\ [1.5 ex]
{\frac{t^4}{t_0^4}}{\bar N}_t^4{\varrho}_{\rm m} & \text{for the 
electromagnetic field,}
\end{array}
\right.
\ena
between the quantity ${\bar {\varrho}}_{\rm m}$ and the directly measurable
passive (inertial) mass density ${\varrho}_{\rm m}$. Now, from equation (26)
we find that
\eqa
{\cal L}_{{\bf {\bar n}}_t}T_{(t){\bar {\perp}}{\bar {\perp}}}=
{\frac{{\bar N}_{t,{\bar {\perp}}}}{{\bar N}_t}}{\Big (}
T_{(t){\bar {\perp}}{\bar {\perp}}}+{\hat T}_{(t)s}^s{\Big )}=
{\frac{t_0^2}{t^2}}{\frac{{\bar N}_{t,{\bar {\perp}}}}{{\bar N}_t^3}}
{\Big (}{\bar {\varrho}}_{\rm m}c^2+3{\bar p}{\Big )},
\ena
while taking the Lie derivative directly of equation (31) we find
\eqa
{\cal L}_{{\bf {\bar n}}_t}T_{(t){\bar {\perp}}{\bar {\perp}}}=
2{\frac{{\bar N}_{t,{\bar {\perp}}}}{{\bar N}_t}}
T_{(t){\bar {\perp}}{\bar {\perp}}}=
2{\frac{t_0^2}{t^2}}{\frac{{\bar N}_{t,{\bar {\perp}}}}{{\bar N}_t^3}}
{\bar {\varrho}}_{\rm m}c^2.
\ena
But then, to be consistent equations (33) and (34) imply that the perfect 
fluid must satisfy the equation of state ${\varrho}_{\rm m}=3p/c^2$, i.e., it 
must be a null fluid. That is, any material component of the fluid must be 
ultrarelativistic, so that any deviation from said equation of state is 
negligible. This is a good approximation for a hot plasma mainly consisting of 
photons and neutrinos. The above result was arrived at also for the OQG, see 
[3] for a further discussion. Note that the null fluid condition follows
from the requirement of isotropy. This means that the assumption of (exact)
isotropy will no longer hold when the cosmic fluid has cooled so much that the 
energy density of non-relativistic particles becomes comparable to that of the 
photons (plus the neutrinos). Rather, cosmologically induced flows will be set 
up and (spatial) metric fluctuations must necessarily occur as seeds for 
structure formation. The details of this and if the resulting predictions are 
consistent with observations is a subject for further work.

It turns out that for the metric family (29) it is sufficient to solve equation
(8) in order to find a solution ${\bar N}_t$ (since equation (20) is 
identically fulfilled). Such a solution with a correct vacuum limit was found 
in [3]. The solution found from equation (8) is given by [3] (expressed by the 
``critical'' density 
${\bar {\varrho}}_{\rm m}^{\rm cr}(t_0){\equiv}{\frac{3}{8{\pi}t_0^2G^{\rm S}}}$)
\eqa
{\bar N}_t={\exp}{\Big [}-{\frac{1}{2}}
{\frac{(x^0)^2}{(ct)^2}}{\frac{{\topstar{\bar {\varrho}}}_{\rm m}(t)}
{{\bar {\varrho}}^{\rm cr}_{\rm m}(t_0)}}
+{\frac{1}{2}}{\frac{{\topstar{\bar {\varrho}}}_{\rm m}(t_1)}
{{\bar {\varrho}}^{\rm cr}_{\rm m}(t_0)}}{\Big ]}, \qquad
{\topstar{\bar {\varrho}}}_{\rm m}{\equiv}{\frac{G^{\rm B}}{G^{\rm S}}}
{\bar {\varrho}}^{\rm (EM)}_{\rm m}+{\bar {\varrho}}^{\rm (MA)}_{\rm m}.
\ena
Here, the epoch $t_1$ is interpreted as the epoch when matter creation
ceases, so that essentially ${\frac{\partial}{{\partial}t}}
{\topstar{{\bar {\varrho}}}}_{\rm m}=0$ for $t{\geq}t_1$. Note that the 
explicit dependence on $t$ of ${\bar N}_t$ shown in equation (35) is not 
determined from the field equations. Rather, this particular dependence was 
chosen for physical reasons. For a further discussion of the solution (35), 
see [3].
\subsection{Weak gravitational waves in vacuum}
If we ignore the global curvature of space, the weak-field (linear) 
approximation of the field equation (20) (or (23)) for vacuum is the same as 
for GR provided that both ${\bf {\bar a}}_{\cal F}$ and ${\bar K}_t$ vanish 
identically. (This can be easily seen directly from equations (14) and (20).) 
Thus the counterpart weak-field GR-solution of locally plane-fronted waves with
two independent polarizations will also be an approximate solution of equation 
(20). The only difference from the GR-solution is that the global cosmic 
expansion is included via the scale factor. Thus the family of line elements 
takes the form
\eqa
{\overline {ds}}_t^2=-(dx^0)^2+{\frac{t^2}{t_0^2}}
{\Big [}(E_{ks}+{\bar {\varepsilon}}_{(t)ks})dx^kdx^s{\Big ]},
\ena
where $E_{ks}dx^kdx^s$ denotes the metric of Euclidean space and where the terms
\eqa
{\bar {\varepsilon}}_{(t)ks}={\Re}[{\bar {\cal A}}_{(t)ks}
{\exp}(i{\bar {\vartheta}}_t)],
\ena
describe the plane wave perturbation from the Euclidean background. Moreover,
we have that
\eqa
{\bar {\vartheta}}_t{\equiv}{\bar k}_{(t)0}(x^0-x^0_1)+{\bar k}_{(t)i}x^i, 
\quad {\bar k}_{(t){\mu}}={\bar {\vartheta}}_{t},_{\mu}, \quad
{\bar k}_{(t)0},_t=-{\frac{1}{t}}{\bar k}_{(t)0}, \quad 
{\bar k}_{(t)j},_t=0,
\ena
where ${\bar {\vartheta}}_t$ is the phase factor and where ${\bar k}_{(t){\mu}}$ 
denotes the components of the wave 4-vector family. (Also, $x^0_1$ is an 
arbitrary reference epoch.) Finally, 
${\bar {\cal A}}_{(t)ks}={\frac{t_0}{t}}{\bar {\cal A}}_{(t_0)ks}$ is the 
(possibly complex) polarization tensor. As for the counterpart GR case, 
equation (23) for vacuum (ignoring global space curvature) yields that the 
plane wave is null, transverse and traceless. That is, choosing Cartesian
coordinates $(x,y,z)$ with the wave travelling in the $z$-direction, equation 
(36) takes the form
\eqa
{\overline {ds}}_t^2=-(dx^0)^2+{\frac{t^2}{t_0^2}}{\Big [}
(1+{\bar {\varepsilon}}_{(t)xx})dx^2+(1-{\bar {\varepsilon}}_{(t)xx})dy^2
+2{\bar {\varepsilon}}_{(t)xy}dxdy+dz^2{\Big ]}.
\ena
Since equations (36) and (39) only describe approximate solutions of equation
(23), to further investigate the nature of gravitational radiation in 
quasi-metric gravity some exact solutions should be found. Such solutions are 
expected to differ from their GR counterparts. However, finding exact wave-like
solutions of equation (23) may turn out to be difficult, and is beyond the 
scope of the present paper.
\section{Conclusion}
In this paper, we have relaxed the original restrictions on the quasi-metric 
space-time geometry $({\cal N},{\bf {\bar g}}_t)$ so that its most general form
is now given by equation (1). The reason for this revision was to make possible
the prediction of (weak) GR-like gravitational waves since such have now been 
directly detected. However, we have shown that it is not possible to construct
a quasi-metric gravitational theory where space-time curvature is fully coupled
to the active stress-energy tensor family ${\bf T}_t$, and such that the 
resulting field equations will have a sensible Newtonian limit. Nevertheless,
we have also shown that the original quasi-metric gravitational field equations
can be extended with the extra equation (20) (or equivalently, one of equations
(17) or (23)) not being fully coupled to ${\bf T}_t$, such that there are no 
obvious problems in the weak-field limit. The extended field equations are, by 
construction, sufficiently flexible and designed to predict GR-like 
gravitational waves in vacuum for the weak-field (linear) approximation. On 
the other hand, exact gravitational wave solutions are expected to differ from 
their GR counterparts. 

Besides the prediction of gravitational waves, the differences between the 
predictions of the extended quasi-metric gravitational theory and the OQG are 
small. In particular, several observations indicating that the cosmic expansion
is relevant for the solar system (constituting a powerful {\em empirical} 
motivation for introducing the QMF in the first place) have identical 
explanations coming from the OQG and the extended theory (see [6] and 
references therein). This means that, disregarding gravitational waves and 
systems emitting gravitational waves (e.g., binary pulsars), the observational 
status of the extended gravitational theory is the same as for the OQG (i.e., 
currently nonviable [3]).
\\ [4mm]
{\bf References} \\ [1mm]
{\bf [1]} D. {\O}stvang, {\em Grav. {\&} Cosmol.} {\bf 11}, 205
(2005) (gr-qc/0112025). \\
{\bf [2]} D. {\O}stvang, {\em Doctoral thesis}, (2001) (gr-qc/0111110). \\
{\bf [3]} D. {\O}stvang, {\em Indian Journal of Physics}, {\bf 92}, 669
(2018) (arXiv:1701.09151). \\
{\bf [4]} K. Kucha{\v{r}}, {\em Journ. Math. Phys.} {\bf 17}, 792 (1976). \\
{\bf [5]} D. {\O}stvang, {\em Grav. {\&} Cosmol.} {\bf 12} 262 (2006) 
(gr-qc/0303107). \\
{\bf [6]} D. {\O}stvang, {\em Grav. {\&} Cosmol.} {\bf 13}, 1 (2007)
(gr-qc/0201097).
\end{document}